\documentclass[useAMS,usenatbib]{mn2e}

\usepackage{ifthen}
\usepackage{epsfig}
\newcounter{Starfound}
\newcommand{\Star}[1]{%
  \setcounter{Starfound}{0}%
  \ifthenelse{\equal{#1}{0043}}{\setcounter{Starfound}{1}XMM~J\thinspace004315.5+412440}{}%
  \ifthenelse{\equal{#1}{0929}}{\setcounter{Starfound}{1}XTE~J\thinspace0929-314}{}%
  \ifthenelse{\equal{#1}{1543}}{\setcounter{Starfound}{1}4U~1543-475}{}%
  \ifthenelse{\equal{#1}{1550}}{\setcounter{Starfound}{1}XTE~J\thinspace1550-564}{}%
  \ifthenelse{\equal{#1}{1655}}{\setcounter{Starfound}{1}GRO~J\thinspace1655-40}{}%
  \ifthenelse{\equal{#1}{1705}}{\setcounter{Starfound}{1}4U~1705-44}{}%
  \ifthenelse{\equal{#1}{1744}}{\setcounter{Starfound}{1}GRO~J\thinspace1744-28}{}%
  \ifthenelse{\equal{#1}{1751}}{\setcounter{Starfound}{1}XTE~J\thinspace1751-305}{}%
  \ifthenelse{\equal{#1}{1807}}{\setcounter{Starfound}{1}XTE~J\thinspace1807-294}{}%
  \ifthenelse{\equal{#1}{1808}}{\setcounter{Starfound}{1}SAX~J\thinspace1808.4-3658}{}%
  \ifthenelse{\equal{#1}{GX339-4}}{\setcounter{Starfound}{1}GX~339-4}{}%
  \ifthenelse{\equal{\arabic{Starfound}}{0}}{\message{Unknown Star: #1}Star:``#1''}{}%
}

\title[Mass transfer during low mass X-ray transient decays]{Mass transfer during low mass X-ray transient decays}
\author[C. R. Powell, C. A. Haswell and  M. Falanga]{%
Craig R. Powell$^{1,2}$\thanks{E-mail: craig.powell@manchester.ac.uk~(CRP);
                                 c.a.haswell@open.ac.uk (CAH)}, 
Carole A. Haswell$^{1}\footnotemark[1]$,
and Maurizio Falanga$^{3}$\\
$^{1}$Department of Physics and Astronomy, The Open University, Walton Hall, Milton Keynes MK7 6AA\\
$^{2}$School of Physics and Astronomy, The University of Manchester, Manchester M13 9PL\\
$^{3}$Service d'Astrophysique, DAPNIA/DSM/CEA, 91191 Gif-sur-Yvette, France}
\begin{document}

\date{in original form 2006 April 13}

\pagerange{\pageref{firstpage}--\pageref{lastpage}} \pubyear{2006}

\maketitle

\label{firstpage}

\begin{abstract}
The outbursts of low mass X-ray binaries are prolonged relative to
those of dwarf nova cataclysmic variables as a consequence of X-ray
irradiation of the disc.  We show that the time-scale of the decay
light curve and its luminosity at a characteristic time are linked to
the radius of the accretion disc.  Hence a good X-ray light curve
permits two independent estimates of the disc radius.  In the case of
the milli-second pulsars \Star{1808} and \Star{0929} the
agreement between these estimates is very strong.  Our analysis allows
new determinations of distances and accretion disc radii.  Our
analysis will allow determination of accretion disc radii for sources
in external galaxies, and hence constrain system parameters where
other observational techniques are not possible.  We also use the
X-ray light curves to estimate the mass transfer rate.  The broken
exponential decay observed in the 2002 outburst of \Star{1808} may be
caused by the changing self-shadowing of the disc.
\end{abstract}


\section{Introduction}\label{Sec:intro}
\citet{KingRitter98} (henceforth KR) and Shahbaz, Char\-les \& King
(1998) (henceforth SCK) examined the X-ray light curves of transient
low mass X-ray binaries (LMXBs) in decay, characterising them as
simple exponential or linear decays.  In some cases, however, the
decays show a distinct knee.  As shown in Fig. \ref{Fig:1808}, this is
not a secondary maximum in the sense that the X-ray count rate
declines throughout.

In X-ray bright disc-accreting systems, including transient LMXBs
during outburst, the temperature of the accretion disc is dominated by
X-ray heating from the inner accretion regions across most of the
disc.  For a disc in which the scale height can be described as a
function of radius by $H=H_0R^n$ for constant $H_0$ and $n$, the
temperature, $T$, is given by \citet{JongParadAugus96} as
\begin{equation}
  T^4=
  \frac{1-\mathcal{A}}{4\pi\sigma R^2}
  \frac{H}{R}\left(n-1\right)
  L_\textrm{\sc x},  
  \label{Eq:TofR}
\end{equation}
where $\mathcal{A}$ is the disc albedo, $L_\textrm{\sc x}$ is the
central X-ray luminosity heating the disc, assumed to be emanating
radially from a small spherical source.  Only if $n>1$ is the whole
surface of the accretion disc illuminated; if $n=1$ (\ref{Eq:TofR})
predicts a surface temperature of zero which is invalid because of
local viscous heat production.  Consistent with KR we adopt a value of
$n=9/7$, although see section~\ref{Sec:dM2small}.  (\ref{Eq:TofR}) can
be rewritten to give the maximum radius $R_h$ which is heated by
X-rays to the temperature $T_h$ needed to remain in the hot, viscous
state.  Thus
\begin{equation}
  R_h^{3-n}=
  \frac{1-\mathcal{A}}{4\pi\sigma T_h^4}
  H_0\left(n-1\right)
  L_\textrm{\sc x}.
  \label{Eq:Rhot}
\end{equation}
We abbreviate this equation as
\begin{equation}
  R_h^{3-n}=\Phi L_\textrm{\sc x}.
  \label{Eq:RhotL}
\end{equation}
We note that especially in the case of a black hole the X-ray source
will be the inner accretion disc, and as such disc-like rather than
point-like.  This introduces an factor of $H/R$ into the r.h.s. of
(\ref{Eq:Rhot}), which is equivalent to modified values of $n$ and
$\Phi$ in (\ref{Eq:RhotL}); specifically, the index of $3-n$ in the
l.h.s. becomes $4-2n$, and $\Phi$ gains a factor of $H_0$.  However,
$n$ and $H_0$ may differ systematically between neutron star and black
hole systems, in which case the differences in (\ref{Eq:RhotL}) will
be less simple.

KR showed that X-ray heating during the decay from outburst causes the
light curves of transient LMXBs to exhibit either exponential or linear
declines depending on whether or not the luminosity is sufficient to
keep the outer disc edge hot.  They note that exponential decays must
revert to the linear mode when the X-ray flux has decreased
sufficiently, but do not analyse this in detail.  Nor do they consider
the effects of mass transfer from the donor star, $-\dot M_2$, during
the outburst.  Whilst in many systems $-\dot M_2$ is negligible
compared to the mass accretion rate onto the compact object, $\dot
M_c$, during outburst, this is not necessarily the case.  In this
paper we consider these issues.  Section~\ref{Sec:theory} contains our
theoretical predictions for the decay light curve.  In
section~\ref{Sec:fitting} we identify some suitable light curves and
describe their parametrization.  We compare the results from the
fitting in section~\ref{Sec:comparison}.

\section{Analysis}\label{Sec:theory}
\subsection{Exponential decay}
The mass of the accretion disc during the exponential decay is given
by equation 3 of KR;
\begin{equation}
  M_{\rm disc}=\frac{\dot M_cR_{\rm disc}^2}{3\nu_{\rm{KR}}},
  \label{Eq:Mdisc}
\end{equation}
where $\nu_{\rm{KR}}$ is some measure of viscosity, taken by KR to be
the viscosity near the outer disc edge with a value of
$\nu_{\rm{KR}}\sim10^{11}{\rm \,m}^2{\rm \,s}^{-1}$.  In the case of
significant mass transfer into the disc from the donor star the
central accretion rate is given by
\begin{equation}
  \dot M_c=-\dot M_2-\dot M_{\rm disc}.
\end{equation}
Using this to replace $\dot M_c$ in (\ref{Eq:Mdisc}) we obtain
\begin{equation}
  -\dot M_2-\dot M_{\rm disc}=\frac{3\nu_{\rm{KR}}M_{\rm disc}}{R_{\rm disc}^2}.
\end{equation}
Therefore the mass of the hot disc can be written
\begin{equation}
  M_{\rm disc}=
  M_\alpha\exp\left(-\frac{3\nu_{\rm{KR}}t}{R_{\rm disc}^2}\right)
  +\frac{R_{\rm disc}^2\left(-\dot M_2\right)}{3\nu_{\rm{KR}}},
\end{equation}
where $M_\alpha$ is the constant of integration,
and the X-ray luminosity is proportional to
\begin{equation}
  \dot M_c=
  -\dot M_2
  +\frac{3M_\alpha\nu_{\rm{KR}}}{R_{\rm disc}^2}\exp\left(-\frac{3\nu_{\rm{KR}}t}{R_{\rm disc}^2}\right).
\end{equation}
At some time, $t_t$, the temperature of the outer disc edge, while
still dominated by X-ray heating, will be only just sufficient to
remain in the hot, viscous state, i.e. $R_h=R_{\rm disc}$.  We denote
the corresponding X-ray luminosity as $L_t$, with the central
accretion rate $\dot M_t$.  The X-ray luminosity at earlier times is
\begin{equation}
  L_\textrm{\sc x}=
  \left(L_t-L_2\right)\exp\left(-\frac{3\nu_{\rm{KR}}\left(t-t_t\right)}{R_{\rm disc}^2}\right)
  +L_2,
  \label{Eq:Roftau}
\end{equation}
where $L_2=\eta\left(-\dot M_2\right)c^2$, and $\eta\simeq0.1$ is the
efficiency with which rest mass is liberated as X-rays.  

\subsection{Linear decay}\label{Sec:linear}
When $R_h<R_{\rm disc}$ the outer part of the disc is no longer kept
in the hot, viscous state by central irradiation.  The radius of the
hot part of the disc will decrease with the decreasing X-ray
luminosity, corresponding to the late linear decline of KR.  This,
too, may be modified by the donor star mass loss rate, depending on
the mass accretion rate through the now cold outer disc.  Designating
$\mu_c\!\left(R\right)$ as the rate at which mass is transported
inwards through the cold disc, we examine the two extreme behaviours
of $\mu_c$.  For the minimally efficient cold disc
$\mu_c\!\left(R\right)=0$ for all $R$, and none of $-\dot M_2$ reaches
the hot disc when $R_h<R_{\rm disc}$.  At the other extreme
$\mu_c\!\left(R\right)=-\dot M_2$ and the cold disc does not increase
in surface density with time.  The occurrence of outbursts suggests
that the latter case is false; with stable cold-state mass transfer,
outbursts would never occur.

The mass in the hot disc changes due to three factors: accretion
from its inner edge, the loss of material into the encroaching
cold disc, and mass transfer from the cold disc.  Hence,
\begin{equation}
  \dot M_{\rm hot}=-\dot M_c+\mu_c\!\left(R_h\right)+2\pi R_h\Sigma\!\left(R_h\right)\dot R_h.
  \label{Eq:Mhotlin}
\end{equation}
Using the approximate substitution from (\ref{Eq:RhotL}) that
\begin{equation}
  R_h^2=\Phi\eta c^2\dot M_c, 
\end{equation}
(\ref{Eq:Mhotlin}) becomes
\begin{equation}
  \dot M_{\rm hot}=-\dot M_c+\mu_c\!\left(R_h\right)+\pi\Phi\eta c^2\Sigma\!\left(R_h\right)\ddot M_c.
  \label{Eq:dMhotlin}
  \end{equation}
Additionally, equation 2 of KR allows $M_{\rm hot}$ to be written as
\begin{equation}
  M_{\rm hot}=\frac{\dot M_cR_h^2}{3\nu_{\rm{KR}}}.
  \label{Eq:KR2}
\end{equation}
Equating (\ref{Eq:dMhotlin}) to ${\rm d}/{\rm d}t$
(\ref{Eq:KR2}),
\begin{equation}
  \frac{1}{3\nu_{\rm{KR}}}\frac{\rm d}{{\rm d}t}\dot M_cR_h^2=
  -\dot M_c+\mu_c\!\left(R_h\right)+\pi\Phi\eta c^2\Sigma\!\left(R_h\right)\ddot M_c,
\end{equation}
so using equation 2 of KR,
\begin{equation}
  \ddot M_c=
  \frac{3\nu_{\rm{KR}}}{\Phi\eta c^2}\left(\frac{\mu_c\!\left(R_h\right)}{\dot M_c}-1\right).
  \label{Eq:ddMcGeneric}
\end{equation}
We now substitute the two cases of $-\mu_c\!\left(R_h\right)$
outlined above.  When mass transfer into the hot disc is negligible,
\begin{equation}
  \ddot M_c=
  -\frac{3\nu_{\rm{KR}}}{\Phi\eta c^2},
  \label{Eq:ddMc}
\end{equation}
and the decay is the linear decline predicted by KR
as expected.  When $\mu_c\!\left(R_h\right)=-\dot M_2$,
\begin{equation}
  \ddot M_c=
  \frac{3\nu_{\rm{KR}}}{\Phi\eta c^2}\frac{\left(-\dot M_2-\dot M_c\right)}{\dot M_c}.
  \label{Eq:ddMc2}
\end{equation}
While $\dot M_c\gg-\dot M_2$ this reduces to the same form.  As $\dot
M_c$ decreases it will approach the limit of $\dot M_c=-\dot M_2$, at
which point $\ddot M_c=0$.  The exact form of the decay will depend on
the unreliably known form of $\mu_c\!\left(R\right)$, and during the
late linear decline viscous heating near $R_h$ becomes
non-negligible.  We do not predict the form of the late decay other
than to note that if there is some radius $R_{\rm lim}$ satisfying
\begin{equation}
  \mu\!\left(R_{\rm lim}\right)\eta c^2=\frac{R_{\rm lim}^2}{\Phi},
  \label{Eq:Lqui}
\end{equation}
then the l.h.s. of (\ref{Eq:Lqui}) will give the asymptotic
limit of the decay and the accretion luminosity at the beginning of
quiescence.

As a final point on the linear decay we note that the viscous
instability may cause a range of annuli within the cold disc to enter
the hot state and rapidly transfer mass inward.  Some of this may
enter the inner hot disc, or else may increase the surface density of
the subsequent cold disc such that the inner hot disc receives more
material at its outer edge.  Consequently, small rebrightenings or
other artefacts potentially complicate the linear decay.

\subsection{Continuous derivative}\label{Sec:cont}
To examine the observed X-ray light curves of transient LMXBs it
is useful to examine the constraints on the transition.  The gradient
of the exponential decay at time $t_t$ is
\begin{equation}
  \dot L_\textrm{\sc x}=
  -\frac{3\nu_{\rm{KR}}}{R_{\rm disc}^2}\left(L_t-\eta\left(-\dot M_2\right)c^2\right),
  \label{Eq:dLxe}
\end{equation}
whilst the gradient of the linear decline is given by (\ref{Eq:ddMc}),
\begin{equation}
  \dot L_\textrm{\sc x}=
  -\frac{3\nu_{\rm{KR}}}{\Phi}.
\end{equation}
From (\ref{Eq:RhotL}) we adopt
\begin{equation}
  R_{\rm disc}^2=\Phi L_t,
  \label{Eq:RdiscofLt}
\end{equation}
where it is assumed that at the outer disc edge
$H\simeq0.2R_{\rm disc}$.  We therefore obtain
\begin{equation}
  \dot L_\textrm{\sc x}\!\left(t_t\right)=
  -\frac{3\nu_{\rm{KR}}}{\Phi}\left(1-\frac{\eta\left(-\dot M_2\right)c^2}{L_t}\right).
\end{equation}
In the case that $-\eta\dot M_2c^2$ is small relative to $L_\textrm{\sc x}$
the gradient of the exponential decay at time $t_t$ is equal to that
of the subsequent linear decline; the first derivative of the X-ray
light curve is expected to vary smoothly throughout the decay.  If
however $-\eta\dot M_2c^2$ cannot be neglected then the linear decline
is steeper than the exponential decay at $t_t$ and a discontinuity in
gradient is expected.  Whether this is detectable depends on the value
of $-\dot M_2$ and the quality of the X-ray data.

Some X-ray light curves, including most of those discussed in this
paper, include a knee feature in the decay from outburst.  Where this
can be explained by an exponential to linear transition with
non-negligible $-\eta\dot M_2c^2$ we label it as a `brink'.  In
general the decay light curve may contain multiple knees due to
different effects, but we do not expect more than one brink unless the
source rebrightens between them such that each occurs at approximately
the same luminosity, corresponding to a disc not varying greatly in
radius.

\section{Application to observed X-ray transients}
\label{Sec:fitting}
X-ray light curves from the All-Sky Monitor (ASM) and Proportional
Counter Array (PCA) of the {\it RXTE} space telescope were examined to
find examples of decay light curves which fit the template predicted
above.  We use our analysis to determine distances and accretion disc
radii.  System parameters from the literature for the objects analysed
are collected in Table~\ref{Tab:syspar} for comparison with quantities
we derive.  Since we will be calculating the radius of the accretion
disc we calculate from the system parameters the orbital separation
$a$.

The count rates $N$ as a function of time $t$ were fitted with
\begin{equation}
  N=(N_t-N_e)\exp\left(-\frac{t-t_t}{\tau_e}\right)+N_e
\end{equation}
for the exponential phase, where $N_e$ is the limit of the exponential
decay, $N_t$ is the count rate at the brink, and $\tau_e$ is the
time-scale of the decay.  The linear decline is given by
\begin{equation}
  N=N_t\left(1-\frac{t-t_t}{\tau_l}\right),
\end{equation}
where $\tau_l$ is the time after the brink at which the count rate
would become zero if the linear decline continued.  Using the assumed
distance to each source with a hydrogen column density of
$10^{22}{\rm\;cm}^{-2}$ it is possible to convert the count rates
$N_t$ and $N_e$ into absolute X-ray luminosities $L_t$ and $L_e$, the
X-ray luminosity at the brink and the limit of the exponential decay
(corresponding to $-\dot M_2$) respectively.  For the two systems for
which we have simultaneous ASM and PCA light curves, \Star{0929} and
\Star{1744}, the ratio of count rates is approximately 31.  We assume
this conversion to be valid for all sources for comparison purposes,
and present luminosities in the 1.5--12$\;{\rm keV}$ range
corresponding to the ASM spectral range using an assumed power law
spectrum with photon index of $\Gamma=2$.  The luminosities thus
determined are given in Table~\ref{Tab:fitpar}.  We note that if the
spectrum differs between systems our deduced $L_t$ is invalid, though
our deduced exponential timescale will be unaffected.

Using (\ref{Eq:RhotL}) and (\ref{Eq:Roftau}) it is possible to
calculate the disc radius independently from the measured values of
$N_t$ and $\tau_e$; the result based on $\tau_e$ is of particular
importance as it does not depend on the conversion of count rate to
flux but only on a directly observable timescale.  It is also possible
to determine $-\dot M_2$ from $N_e$.  Results are given in
Table~\ref{Tab:calcpar}; for comparison we have also listed the
circularization radius, $R_{\rm circ}$, and the distance between the
compact object and the inner Lagrange point, $b_1$, for each source
based on the values in Table~\ref{Tab:syspar}.  Discussion of the
choice of values for $\Phi$ and $\nu_{\rm{KR}}$ is given in
section~\ref{Sec:comparison}.

\subsection{\protect\Star{1543}}
\begin{figure}
  \epsfig{angle=0,width=84mm,file=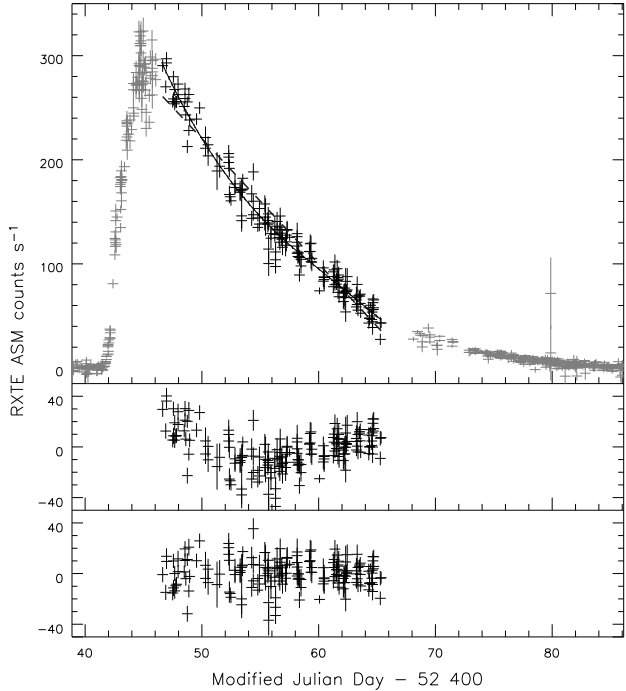}
  \caption{{\bf Top panel:} The fitted ASM light curve of
           4U\thinspace1543-475.  The solid line shows the
           exponential-then-linear fit, the dashed line indicates the
           best linear fit.  The data shown in grey were not included
           in the fit, lying before and after the period of interest.
           {\bf Middle panel:} Residuals to the linear fit.
           {\bf Bottom panel:} Residuals to the exponential/linear fit.
           }
  \label{Fig:1543}
\end{figure}%
Fig.~\ref{Fig:1543} shows that the exponential-then-linear fit
to the decay from the 2002 outburst of \Star{1543} is significantly
better than the least-squares regression line.  Because no
discontinuity is seen in the gradient, the fitting routine used was
based on the continuous derivative model (see section~\ref{Sec:cont}).
Previous outbursts were reported in 1971, 1983 and 1992.  The source
therefore spends only a small fraction of its time in outburst,
suggesting that $-\dot M_2$ is small compared with the central
accretion rate during outburst, consistent with the source's failure
to display a broken gradient.

\subsection{\protect\Star{1808}}
\begin{figure}
  \epsfig{angle=0,width=84mm,file=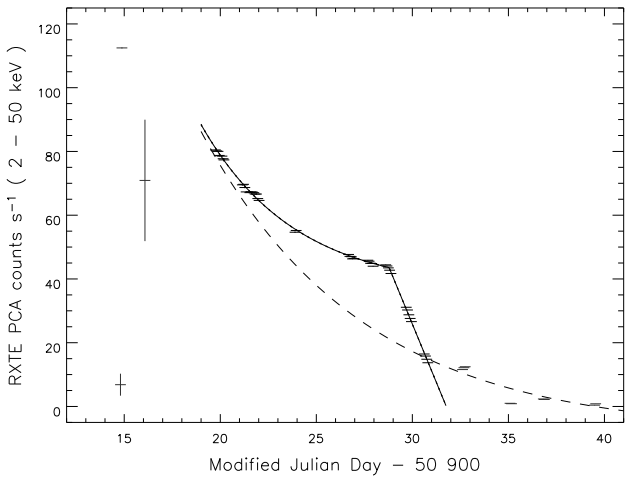}
  \caption{The fitted PCA light curve of the 1998 outburst of
  XTE~J\thinspace1808.4-3658.  The dashed line indicates the fit of SCK.}
  \label{Fig:1808}
\end{figure}%
\begin{figure}
  \epsfig{angle=0,width=84mm,file=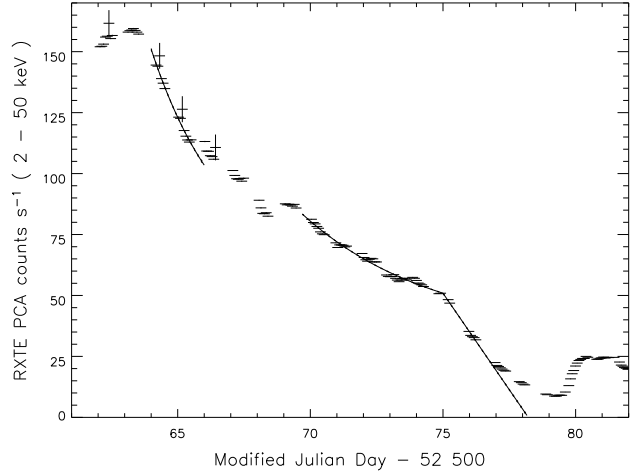}
  \caption{The fitted PCA light curve of the 2002 outburst of
  XTE~J\thinspace1808.4-3658}
  \label{Fig:1808_2002}
\end{figure}%
\Star{1808} was first detected in outburst in 1996 and has
subsequently shown outbursts in 1998, 2000, 2002, and 2005.  Hence it is
expected that the ratio of $-\dot M_2$ to the peak central accretion
rate is much higher than in \Star{1543}.  This is consistent with the
appearance of a broken decay in the two well observed outbursts, shown
in Figs.~\ref{Fig:1808} and \ref{Fig:1808_2002}.  Each light curve is
well fitted by a decay that is exponential towards a positive limit
followed by a sharp gradient discontinuity (brink) leading to a linear
decline.  The fit to the 1998 outburst includes all points that can be
confirmed to occur during the decay; the observation at day 16 of
Fig.~\ref{Fig:1808} may belong to a period of constant luminosity or
even an unsteady rise to outburst, while the observations around day
33 of Fig.~\ref{Fig:1808} may involve a rebrightening.  Even if this
is not the case the late linear decline is expected to become
increasingly complex, as we noted in Section~\ref{Sec:linear}.
Comparison with Fig.~\ref{Fig:1808_2002} shows that the brink occurs
at approximately the same count rate in both cases.  This is expected
if $R_{\rm disc}$ remains constant (i.e. $L_t$ is the same).

The dashed line in Fig.~\ref{Fig:1808} indicates the fit proposed by
SCK, in which they identified the surplus around the brink as a
secondary maximum.  We disagree and view their fit as inadequate in
having a negative limit to the exponential decay.  It is possible that
features in other decay light curves identified as secondary maxima but
having monotonically decreasing X-ray flux have the same explanation.

A second issue in Fig.~\ref{Fig:1808_2002} is the broken nature of the
light curve prior to the brink.  Whilst we do not have a detailed
explanation for this, we speculate that it may be associated with some
regions of the disc being self-shadowed and initially remaining in the
cool state.  3D hydrodynamic simulations \citep[e.g.][]{FoulkesETAL06}
show self-shadowing is likely, due to spiral density waves and
irradiated disc warping \citep[c.f.][]{Pringle96}.  Consequently the
effective area of the hot disc, the quantity probably corresponding to
our measurement of $R_{\rm disc}^2$, increases during the decay in an
apparently stepwise manner, introducing new material into the hot disc
and therefore resetting the central luminosity to a higher value.
This view is supported by the fact that fitting the region around day
65 of Fig.~\ref{Fig:1808_2002} indicates a shorter exponential
time-scale than from day 70 to the brink, corresponding to a smaller
outer radius of the hot region.  By day 70 we assume that the entire
disc is in the hot state. The growing irradiated area could lead to an
actual increase in $L_\textrm{\sc x}$, hence it is an explanation for
a type of secondary maximum.  It is possible that the light curve will
monotonically decline despite the growing irradiated area, hence
producing knees in the light curve.  The light curve of \Star{1808}
around day 67 of Fig.~\ref{Fig:1808_2002} is suggestive of this
behaviour.

\subsection{\protect\Star{1751}}
\begin{figure}
  \epsfig{angle=0,width=84mm,file=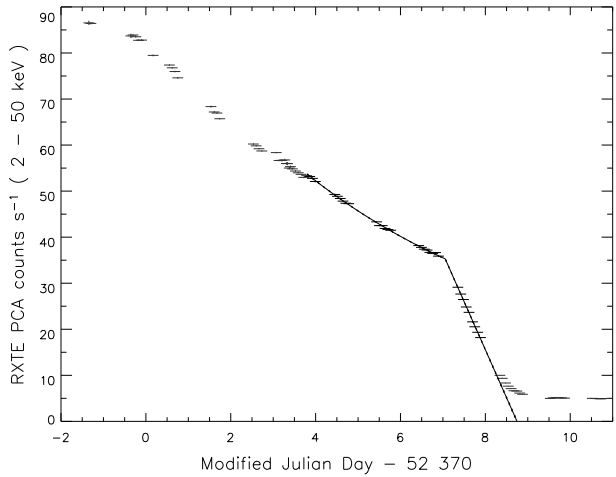}
  \caption{The PCA light curve of the 2002 April outburst of
  XTE~J\thinspace1751-305, fitted after MJD 52 373.}
  \label{Fig:1751}
\end{figure}%
\Star{1751} was first detected during its 2002 outburst, shown in
Fig.~\ref{Fig:1751}, and a second outburst was observed in 2005.  If
this repetition interval is normal for the system, it indicates a
ratio of $-\dot M_2$ to peak central accretion rate more similar to
that of \Star{1808} than to \Star{1543}.  Consistent with this, the
2002 decay light curve has a shape similar to those of \Star{1808} in
that a smooth exponential decay is broken by occasional
rebrightenings.  Therefore only the last portion, appearing to
correspond to a single smooth decay, is fitted.  The linear portion of
this decay terminates at a constant level.  In accordance with
(\ref{Eq:ddMc2}) we interpret this as the rate at which mass can be
transferred through the mostly cold disc; this rate will continue
during quiescence whilst the disc mass increases.

\subsection{\protect\Star{0929}}
\begin{figure}
  \epsfig{angle=0,width=84mm,file=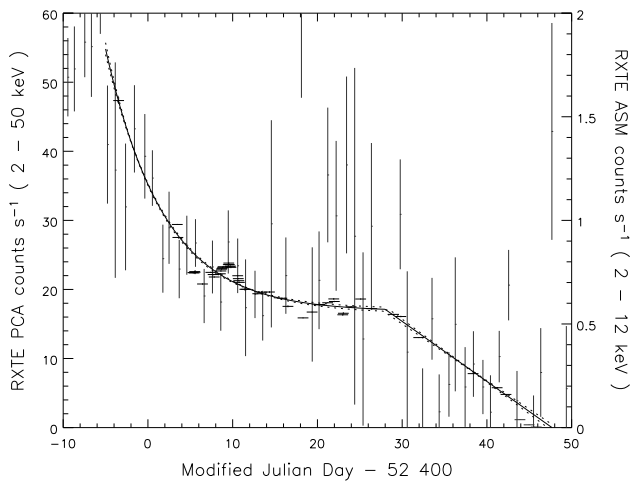}
  \caption{The fitted PCA light curve of XTE J\thinspace0929-314.  The ASM
  light curve, suitably scaled, is overplotted in grey.}
  \label{Fig:0929}
\end{figure}%
A third observed example of the brink is found in the only observed
(2002) outburst of \Star{0929}.  Here the overall shape before the brink
is exponential, but a maximum at about day 10 of Fig.~\ref{Fig:0929}
and a minimum at about day 19 are anomalous.  Possibly the outer edge
of the disc remained cold until late in the exponential decay and
several secondary maxima of the type already discussed have occurred.

\subsection{\protect\Star{1807}}
\begin{figure}
  \epsfig{angle=0,width=84mm,file=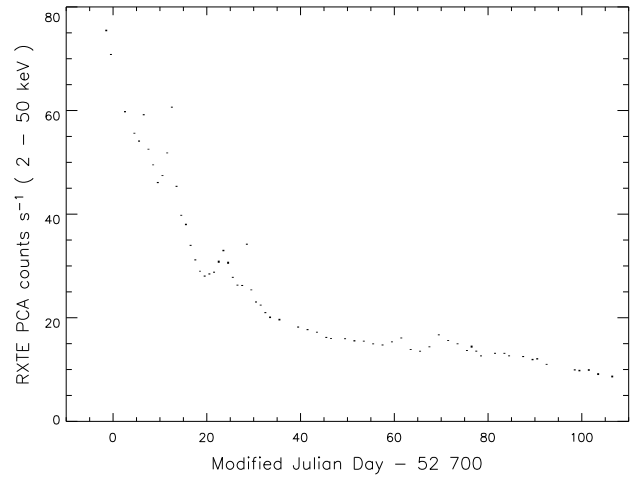}
  \caption{The PCA light curve of XTE~J\thinspace1807-294}
  \label{Fig:1807}
\end{figure}%
\Star{1807} is, like \Star{1808}, \Star{1751} and \Star{0929}, one of
seven known milli-second pulsar (MSP) systems.  MSPs have short
orbital periods and hence small and therefore relatively simple discs.
We therefore expect the X-ray light curve of \Star{1807} to also
display the brinked decay seen in the first three.
Fig.~\ref{Fig:1807} does not appear to lend itself to this type of
fitting, and we have been unable to obtain reasonable fitting
parameters.  One possible explanation is that the linear decay had
already begun at the start of the observed light curve; in this case
the brink count rate must be around fifty per cent greater than that
seen in \Star{1808}.  The orbital separation of \Star{1807}
(c.f. Table~\ref{Tab:syspar}) is half that of \Star{1808}, so the disc
must be correspondingly smaller.  A high value of $N_t$ can only be
accounted for if \Star{1807} is at a distance of less than about
$1{\rm\,kpc}$ so the luminosity remains low while the countrate
is high.  Conversely, if the light curve corresponds to the
exponential portion with several rebrightenings, the maximum value of
$N_t$ may be 15, with the light curve after day 80 tentatively assumed
to be linear.  If the disc is half the radius of that of \Star{1808},
in proportion to the orbital separation, (\ref{Eq:dMhotlin}) indicates
that the brink luminosity will be one quarter that of \Star{1808}.
The PCA count rate at the proposed brink is one quarter that of
\Star{1808}, indicating a similar source distance of
$D\simeq2{\rm\,kpc}$.  We deem the latter explanation more likely, and
suspect that the exponential decay of \Star{1807} is complicated by
self-shadowing of the disc.

The more extreme mass ratio of \Star{1807} than \Star{1808}, indicated
in Table~\ref{Tab:syspar}, suggests that the accretion disc radius in
\Star{1807} is a larger fraction of the orbital separation, indicating
a larger brink luminosity than assumed above, increasing the distance
estimates in both cases.  A lower value of $N_t$ is possible in the
second case, which would also indicate a greater source distance.

\subsection{\protect\Star{1744}}
\begin{figure}
  \epsfig{angle=0,width=84mm,file=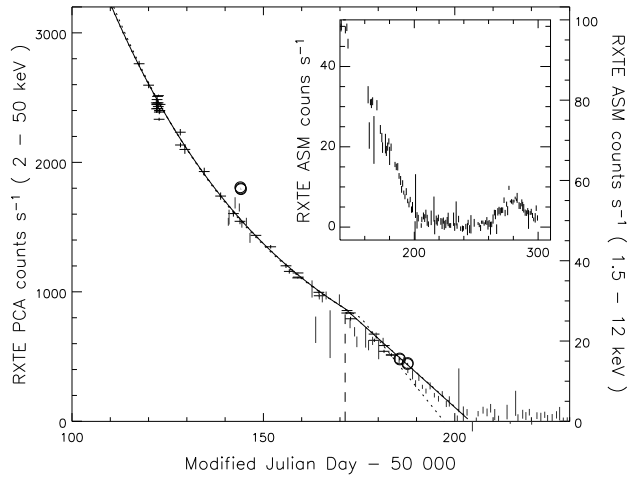}
  \caption{GRO~J\thinspace1744-28 light curve for the decay of 1996.
           Points shown in black are the PCA data; circles indicate
           those data not fitted.  ASM data is shown in grey to mark
           the behaviour during a gap in the PCA coverage.  The solid
           line indicates the best-fitting decay model; the dotted
           lines indicate the error bars on the model.  The dashed
           line marks the best-fitting date of the brink.  Inset: The
           ASM light curve for the late decay and subsequent
           mini-outburst.
          }
  \label{Fig:1744_1996}
\end{figure}%
\begin{figure}
  \epsfig{angle=0,width=84mm,file=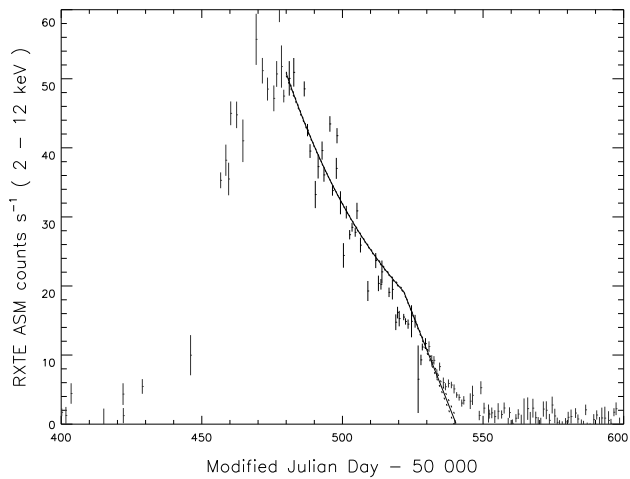}
  \caption{GRO~J\thinspace1744-28 light curve for the decay of 1997.
           Solid and dotted lines mark the best fit and uncertainty,
           as in Fig.~\ref{Fig:1744_1996}.}
  \label{Fig:1744_1997}
\end{figure}%
The earliest part of the {\it RXTE} light curve of \Star{1744}, shown
in Fig.~\ref{Fig:1744_1996}, shows the system already in outburst; the
first PCA data are consistent with the outburst decay.  The light curve
has a brink just after day 170 in Fig.~\ref{Fig:1744_1996}.
\Star{1744} entered a relatively quiescent period in 1996 April
($\sim$day 200 of Fig. \ref{Fig:1744_1996}), punctuated by a
mini-outburst shown inset in Fig.~\ref{Fig:1744_1996} following day
260 which reached less than half the brink luminosity.  A second
outburst, smaller than the 1996 outburst, followed during the first
four months of 1997.  Since then the system has been quiescent.
Fig.~\ref{Fig:1744_1997} shows the decay from the 1997 outburst, in
which a brink is clearly present.  Our confidence in the
identification of these two knees as brinks is increased by their
similarity in luminosity.

\subsection{\protect\Star{GX339-4}}
\begin{figure}
  \epsfig{angle=0,width=84mm,file=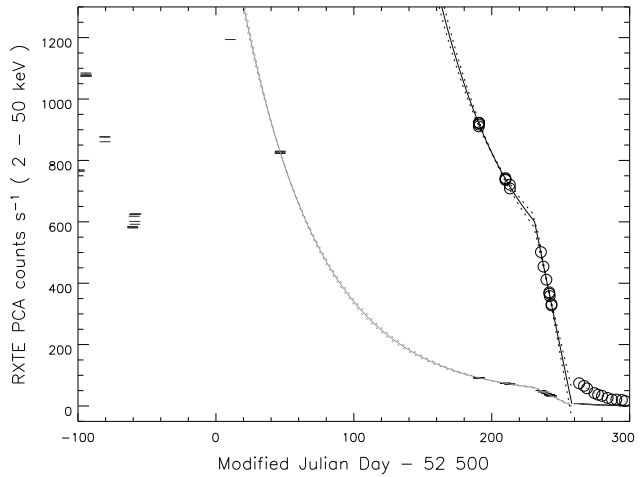}
  \caption{The fitted PCA light curve of GX~339-4.  Circles mark data
  points where the y-scale has been stretched by a factor of ten.}
  \label{Fig:GX339-4}
\end{figure}%
As shown in Fig.~\ref{Fig:GX339-4}, \Star{GX339-4} follows an
exponential decay to less than 10 per cent of its outburst maximum, a
much smaller fraction than is exhibited by the other systems
considered so far.  The relatively few existing points fit the brinked
decay model well.  The late decay departs smoothly from the linear
decline, in line with expectations (c.f. section~\ref{Sec:linear}
eqn.\ \ref{Eq:ddMcGeneric} and following comments).

\subsection{\protect\Star{1550}}
\begin{figure}
  \epsfig{angle=0,width=84mm,file=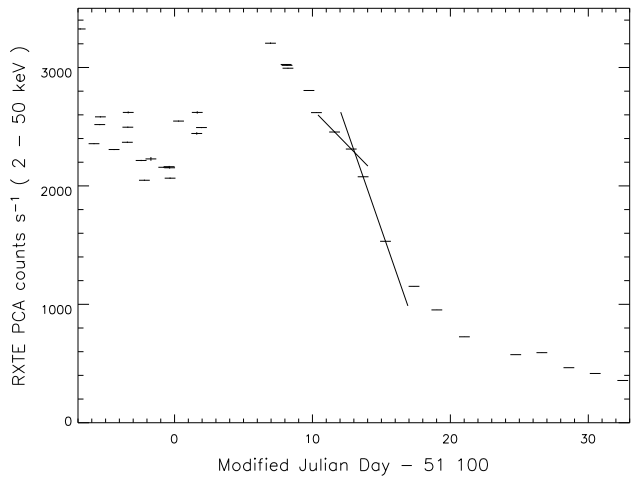}
  \caption{The PCA light curve of XTE~J\thinspace1550-564.  Not enough
  data cover the decay to be able to be able to perform useful
  fitting, but the presence of a knee in the decay is indicated by the
  four points through which lines have been drawn.}
  \label{Fig:1550}
\end{figure}%
Although not enough data exist to be able to establish the parameters
of the brinked decay, a knee occurs in the PCA light curve, which has
been marked in Fig.~\ref{Fig:1550}.  Interpreting this as the brink
it is therefore possible to calculate $R_{\rm disc}$ based on $N_t$
but no other system parameters.

\subsection{\protect\Star{1655}}
\begin{figure}
  \epsfig{angle=0,width=84mm,file=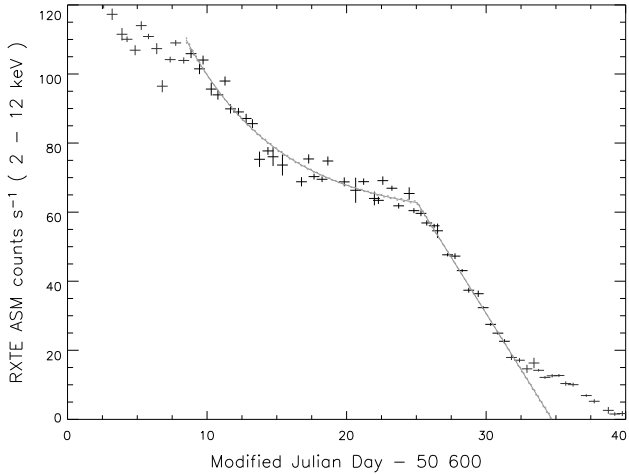}
  \caption{The ASM light curve of GRO~J\thinspace1655-40.  Although the
  calculated error on the brink luminosity is less than
  1 ASM count ${\rm s}^{-1}$, we note by examination that the brink
  in the light curve may be uncertain by 5 ASM counts ${\rm s}^{-1}$
  if some of the scatter is intrinsic variability.}
  \label{Fig:1655}
\end{figure}%
The ASM light curve of \Star{1655} shows one outburst, starting in
1996 April and persisting until 1997 August.  This outburst is broadly
divided into two broad peaks, of which the decay of the second is
shown.  Although the system had been in outburst for a considerable
time, \citet{HynesETAL98} note that no stable hot disc state exists
for luminosities less than the Eddington luminosity; it is therefore
unclear whether material from the outer disc edge participated in the
outburst.  As shown in Table~\ref{Tab:calcpar}, the accretion disc
radius calculated from the brink luminosity and the exponential
timescale match each other, but are considerably less than the system's
circularization radius.  This may indicate that the mechanism examined
here is applicable, but acts on a small inner disc region
rather than on the disc as a whole.  Possibly this second peak is
caused by a high surface density remnant left by the preceding
outburst activity.

\Star{1655} returned to outburst in 2005; if the average flux between
the beginning of the two successive outbursts is taken as an indicator
of $-\dot M_2$, this value is found to lie below the brink by a factor
of three.  The mass transfer rate from the secondary star may be
enhanced during outburst, but it is likely that the mass flow onto the
(constant radius) outer edge of the hot disc during the exponential
phase is supplied from additional disc structures rather than from the
secondary directly.  The complexity of the disc in this system does
not necessarily invalidate the application of this model during at
least some phases of the decay.

\subsection{\protect\Star{1705}}
\begin{figure}
  \epsfig{angle=0,width=84mm,file=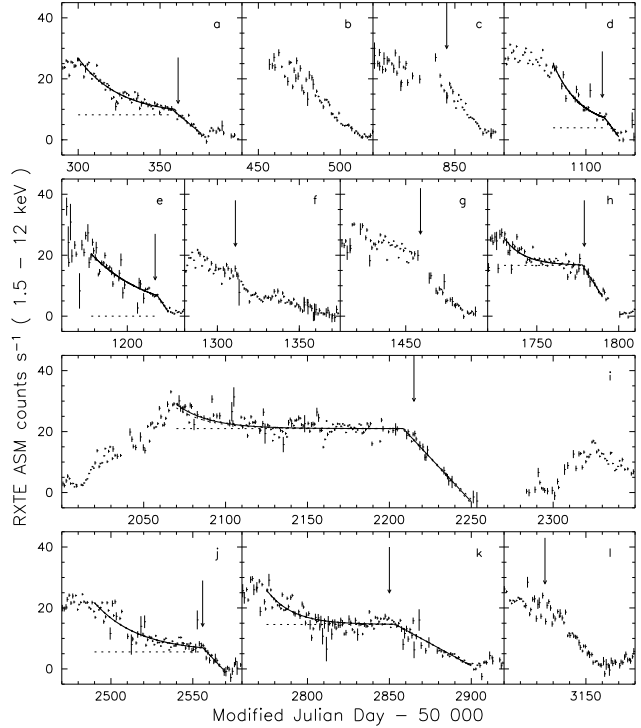}
  \caption{The ASM light curve for all decays from major outburst seen
  in 4U 1705-44 by the instrument to date.  Arrows marks points in the
  light curve where a knee between an exponential decay and a linear
  one may be present, assigned by eye and used as initial conditions
  for the fitting.  Horizontal dotted lines indicate the limit of the
  exponential decay.  Each outburst has been assigned a letter for
  reference.}
  \label{Fig:1705}
\end{figure}%
\begin{figure}
  \epsfig{angle=0,width=84mm,file=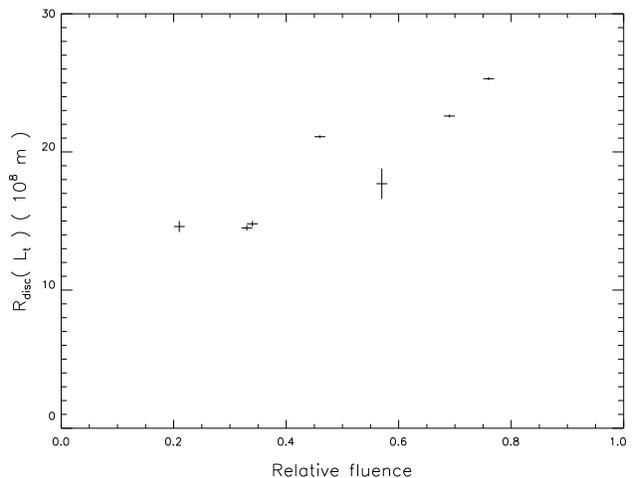}
  \caption{Disc radius versus outburst fluence for the seven fitted
  outbursts of 4U~1705-44.}
  \label{Fig:1705flu}
\end{figure}%
No PCA light curves for \Star{1705} are available, so the best evidence
for our model being applicable to this system comes from the ASM
light curve.  This shows numerous outbursts, the decays of which are
shown in Fig.~\ref{Fig:1705}.  A range of smaller
outbursts also occur which are not shown in Fig.~\ref{Fig:1705}.
Arrows mark features in the light curve that we have interpreted as the
brink of each decay; using these as part of the initial conditions for
finding a best fit we obtained the best-fitting decays overplotted.
Clearly there is substantial variation in the height of the brink,
which indicates variation in the radius of the disc.  In order to
explain this we used the ASM light curve to assess the fluence of each
outburst.  A greater fluence suggests a greater initial disc mass,
which leads us to expect positive correlation between the fluence and
calculated disc radius.  Relative fluences are given in
Table~\ref{Tab:calcpar}, and Fig.~\ref{Fig:1705flu} shows a positive
correlation between outburst fluence and
$R_{\rm{disc}}\!\left(L_t\right)$.

If at least some of the knees marked in Fig.~\ref{Fig:1705} actually
correspond to the disc radius mechanism being considered (i.e. they
are brinks), then substantial variation in the radius is seen.
Furthermore, the limit of the exponential decay is clearly in several
cases below the mean long term count rate of 12 counts per second,
based on the numerical integration of the ASM light curve.  We account
for this by the additional theory given in section~\ref{Sec:dM2small}.

\section{Source data and best-fitting parameters}\label{Sec:comparison}
\begin{table*}
\begin{center}
\begin{tabular}{lccccccc}
\hline\hline
Source &  $P_{\rm orbit}$ [min] & $P_{\rm spin}$ [ms]& $a$ [$10^8$m] &
$M_{2}$ [$M_{\odot}$] & Distance [kpc] & \\ 
\hline
\Star{0929}    &  43.6$^a$      &5.4$^a$ & 3.2       &0.008--0.03$^a$  & 5$^a$       & NS \\
\Star{1543}    &$1617\pm12^b$   &  --    &$73\pm4^c$ &$2.7\pm1.0^c$    &$7.5\pm1.0^c$& BH \\
\Star{1550}    &$2235\pm14^d$   &  --    &86--92$^d$ &$1.4\pm0.5^d$    &2.8--7.6$^d$ & BH \\
\Star{1655}    &$3775.1\pm0.2^e$&  --    &$117\pm1^e$&$2.34\pm0.12^e$  & 3.2$^e$     & BH \\
\Star{1705}    &       --       &  --    &    --     &      --         &$7.3\pm1.0^f$& NS \\
\Star{1744}    &$17000^g$       &  --    &$190^g$    &0.2--0.7$^g$     & 8.5$^h$     & NS \\
\Star{1751}    &$42^i$          &2.3$^i$ & 3.1       &0.013--0.035$^i$ & --          & NS \\
\Star{1807}    &$40^j$          &5.25$^j$& 3.0       &0.01--0.022$^k$  & --          & NS \\
\Star{1808}    &$120.8^l$       &2.5$^l$ & 6.3       &0.04--0.1$^l$    &2.5$^m$      & NS \\
\Star{GX339-4} &$2500^n$        &  --    &80--120$^n$& $\la1.1^n$      & 15$^n$      & BH \\
\hline
\end{tabular}
{\scriptsize
\begin{tabular}{lll}
 a: \citet{GallowayETAL02}     &
 f: \citet{HaberlTitar95}      &
 k: \citet{FalangaETAL05}      \\
 b: \citet{OroszETAL98}        &
 g: \citet{RappaJoss97}        &
 l: \citet{ChakrMorgan98}      \\
 c: \citet{ParkETAL04}         &
 h: \citet{NishiuchiETAL99}    &
 m: \citet{ZandETAL01}         \\
 d: \citet{OroszETAL02}        &
 i: \citet{MarkwardtETAL02}    &
 n: \citet{HynesETAL04}        \\
 e: \citet{OroszBailyn97}      &
 j: \citet{CampanaETAL03}      &
 \\
\end{tabular}
}
\caption{System parameters from the literature.  Orbital separation
  $a$ is based on a $1.4M_{\odot}$ neutron star where an accretor mass is
  not available.}
\label{Tab:syspar}
\end{center}
\end{table*}%
\begin{table*}
\begin{center}
\begin{tabular}{lcccc}
\hline\hline
Decay            & $L_t$ [ $10^{27}{\rm\,J\,s}^{-1}$ ]
                                  & $L_e$ [ $10^{27}{\rm\,J\,s}^{-1}$ ]
                                                 & $\tau_e$ [ days ]
                                                                  & $\tau_l$ [ days ] \\
\hline
\Star{0929}      & $   49\pm1   $ & $  48\pm1  $ & $  6.9\pm0.4$  & $  19.7\pm1.2  $ \\
\Star{1543}      & $27100\pm900 $ &              & $12.05\pm0.13$ &                  \\
\Star{1550}      & $11200\pm100 $ &              &                &                  \\
\Star{1655}      & $ 2900\pm20  $ & $2610\pm50 $ & $ 13.7\pm0.9$  & $  19.3\pm0.3  $ \\
\Star{1705} a    & $ 2400\pm300 $ & $1920\pm80 $ & $ 25.1\pm1.2$  & $  22.7\pm1.4  $ \\
\Star{1705} d    & $ 1680\pm50  $ & $ 900\pm200$ & $ 16.9\pm1.9$  & $  10.0\pm0.6  $ \\
\Star{1705} e    & $ 1650\pm80  $ &  0--220      & $   36\pm3$    & $   8.4\pm0.8  $ \\
\Star{1705} h    & $ 3940\pm40  $ & $3900\pm50 $ & $ 12.0\pm1.4$  & $  19.8\pm1.7  $ \\
\Star{1705} i    & $ 4910\pm20  $ & $4910\pm20 $ & $ 19.7\pm1.0$  & $  36.4\pm1.6  $ \\
\Star{1705} j    & $ 1620\pm40  $ & $4170\pm30 $ & $ 26.8\pm1.4$  & $  12.0\pm1.7  $ \\
\Star{1705} k    & $ 3430\pm20  $ & $3410\pm20 $ & $ 15.9\pm0.7$  & $  50.0\pm0.9  $ \\
\Star{1744} 1996 & $ 8050\pm100 $ &  0--40       & $46.61\pm0.12$ & $  21.4\pm1.0  $ \\
\Star{1744} 1997 & $ 6120\pm80  $ &  0--170      & $ 42.9\pm0.8$  & $  18.7\pm1.2  $ \\
\Star{1751}      & $  361\pm1   $ & $ 110\pm30 $ & $  5.9\pm0.5$  & $  1.69\pm0.02 $ \\
\Star{1808} 1998 & $38.54\pm0.03$ & $32.3\pm0.2$ & $ 4.89\pm0.06$ & $ 2.978\pm0.007$ \\
\Star{1808} 2002a&                & $  50\pm7  $ & $  2.8\pm0.3$  &                  \\
\Star{1808} 2002b& $ 45.1\pm0.1 $ & $29.6\pm0.6$ & $  5.0\pm0.1$  & $  3.22\pm0.03 $ \\
\Star{GX339-4}   & $ 1916\pm90  $ & $ 900\pm200$ & $ 57  \pm3  $  & $    27\pm4    $ \\
\hline
\end{tabular}
\caption{Decay light curve parameters.  Luminosities are based on the
         photon index and distance from Table~\ref{Tab:syspar} and
         values of $N_H$ from the same sources, and are stated for a
         spectral range of 1.5--12$\;{\rm keV}$.  For consistency with
         Figs.~\ref{Fig:1744_1996} and \ref{Fig:0929} we take 1 ASM
         count ${\rm s}^{-1}$ to equal 31 PCA counts ${\rm s}^{-1}$
         for our purposes, suggesting that $\Gamma\simeq2.1$.}
\label{Tab:fitpar} 
\end{center}
\end{table*}%
\begin{table*}
\begin{center}
\begin{tabular}{lcccccc}
\hline\hline
Decay            & $R_{\rm circ}$
                               & $b_1$        & $R_{\rm disc}\!\left(L_t\right)$
                                                                 & $R_{\rm disc}\!\left(\tau_e\right)$
                                                                                   & Relative
                                                                                          & $-\dot M_2$ \\
                 & [ $10^8{\rm\,m}$ ]
                               & [ $10^8{\rm\,m}$ ]
                                              & [ $10^8{\rm\,m}$ ]
                                                                 & [ $10^8{\rm\,m}$ ]
                                                                                   & fluence
                                                                                          & [ $10^{-12}M_{\sun}{\rm\,yr}^{-1}$ ] \\
\hline
\Star{0929}      &  1.5--1.9   &  2.63--2.82  & $ 2.52 \pm0.03 $ & $2.67 \pm0.08 $ &      & $  84  \pm  2  $ \\
\Star{1543}      & 12.7--17.1  &  43.4--48.7  & $18.8  \pm0.3  $ & $3.16 \pm0.02 $ &      &         --       \\ 
\Star{1550}      & 20.0--26.4  &  58.5--64.2  & $12.15 \pm0.05 $ &        --       &      &         --       \\ 
\Star{1655}      & 21.3--22.1  &  70.9--72.1  & $ 6.14 \pm0.02 $ & $3.43 \pm0.11 $ &      & $4600  \pm100  $ \\ 
\Star{1705} a    &     --      &      --      & $17.7  \pm1.1  $ & $5.10 \pm0.12 $ & 0.57 & $3400  \pm200  $ \\ 
\Star{1705} b    &     --      &      --      &         --       &        --       & 0.32 &         --       \\ 
\Star{1705} c    &     --      &      --      &         --       &        --       & 1.00 &         --       \\ 
\Star{1705} d    &     --      &      --      & $14.8  \pm0.2  $ & $4.2  \pm0.2  $ & 0.34 & $1600  \pm400  $ \\ 
\Star{1705} e    &     --      &      --      & $14.6  \pm0.4  $ & $6.1  \pm0.3  $ & 0.21 & $    \la400    $ \\ 
\Star{1705} f    &     --      &      --      &         --       &        --       & 0.21 &         --       \\ 
\Star{1705} g    &     --      &      --      &         --       &        --       & 0.30 &         --       \\ 
\Star{1705} h    &     --      &      --      & $22.6  \pm0.12 $ & $3.5  \pm0.2  $ & 0.69 & $6900  \pm200  $ \\ 
\Star{1705} i    &     --      &      --      & $25.3  \pm0.05 $ & $4.52 \pm0.12 $ & 0.76 & $8600  \pm 50  $ \\ 
\Star{1705} j    &     --      &      --      & $14.5  \pm0.18 $ & $5.27 \pm0.14 $ & 0.33 & $7300  \pm 50  $ \\ 
\Star{1705} k    &     --      &      --      & $21.1  \pm0.06 $ & $4.06 \pm0.09 $ & 0.46 & $6000  \pm 50  $ \\ 
\Star{1705} l    &     --      &      --      &         --       &        --       & 0.29 &         --       \\ 
\Star{1744} 1996 & 32.5--52.3  & 112.1--133.9 & $32.3  \pm0.2  $ & $6.95 \pm0.01 $ &      & $    \la70     $ \\ 
\Star{1744} 1997 & 32.5--52.3  & 112.1--133.9 & $28.2  \pm0.2  $ & $6.67 \pm0.06 $ &      & $   \la300     $ \\ 
\Star{1751}      &  1.4--1.7   &   2.5--2.7   & $ 6.85 \pm0.01 $ & $2.47 \pm0.11 $ &      & $ 200  \pm 50  $ \\ 
\Star{1808} 1998 & 2.06--2.71  &  4.68--5.07  & $ 2.238\pm0.001$ & $2.252\pm0.014$ &      & $  57  \pm  4  $ \\ 
\Star{1808} 2002a& 2.06--2.71  &  4.68--5.07  &         --       & $1.70 \pm0.09 $ &      & $  90  \pm 10  $ \\ 
\Star{1808} 2002b& 2.06--2.71  &  4.68--5.07  & $ 2.421\pm0.003$ & $2.28 \pm0.02 $ &      & $  51.8\pm  1.1$ \\ 
\Star{GX339-4}   &   $>23$     &    $>66$     & $ 4.99 \pm0.12 $ & $6.88 \pm0.23 $ &      & $1600  \pm400  $ \\ 
\hline
\end{tabular}
\caption{System radii and mass transfer rate.  The two values of
         $R_{\rm disc}$ and $-\dot M_2$ are calculated from the X-ray
         light curves; $R_{\rm circ}$ and $b_1$ are calculated from the
         data in Table~\ref{Tab:syspar}.  Errors exclude the effect of
         the uncertainty in source distance.}
\label{Tab:calcpar}
\end{center}
\end{table*}%
To force the disc radii based on $\tau_e$ to lie within the range
$R_{\rm{circ}}<R_{\rm{disc}}\!\left(\tau_e\right)<b$ for the two best
described systems, \Star{1808} and \Star{0929}, the viscosity of
$\nu_{\rm{KR}}=4\times10^{10}{\rm\,m}^2{\rm\,s}^{-1}$ was adopted,
close to the value of $10^{11}{\rm\,m}^2{\rm\,s}^{-1}$ proposed by KR.
To make $R_{\rm{disc}}\!\left(L_t\right)$ consistent with
$R_{\rm{disc}}\!\left(\tau_e\right)$ for most of the neutron star
systems a value of $\Phi=1.3\times10^{-12}{\rm\,m}^2{\rm\,s\,J}^{-1}$
was adopted.  A better fit was found for the black hole systems by
reducing $\Phi$ by a factor of ten, consistent with the extra factor
of $H_0$ discussed in the introduction.  $\Phi$ is somewhat small
compared with the range of
$4<\Phi/(10^{-12}{\rm\,m}^2{\rm\,s\,J}^{-1})<9$ adopted by KR.
Furthermore, since $R_{\rm{disc}}$ is based on $\Phi L_t$, our
systematic underestimate of $L_t$ by considering only the flux in the
1.5--12$\;{\rm keV}$ band suggests a yet smaller value of $\Phi$.
Calculated disc radii using these parameter values are given in
Table~\ref{Tab:calcpar}.

\begin{figure}
  \epsfig{angle=0,width=84mm,file=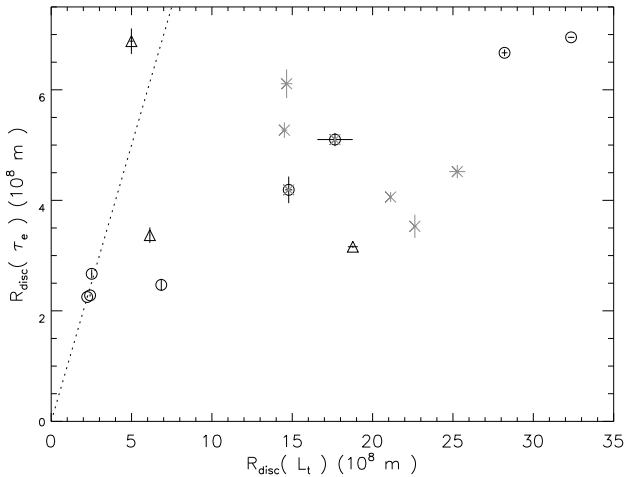}
  \caption{The relation between the radius as determined from the
           brink luminosity against that based on the exponential
           time-scale.  Circles mark neutron star systems, and
           triangles black holes; individual points can be identified
           using Table~\ref{Tab:calcpar}.  Crosses mark decays of
           \protect\Star{1705} other than {\it a} and {\it d}, as
           described in the text.  A diagonal, dotted line marks
           $R_{\rm{disc}}\!\left(L_t\right)=R_{\rm{disc}}\!\left(\tau_e\right)$.
  }
  \label{Fig:radiiratio}
\end{figure}%
Fig.~\ref{Fig:radiiratio} shows the relation between the values of
$R_{\rm disc}$ calculated from $L_t$ and $\tau_e$.  We note that for
\Star{1705} the results would be more consistent if we adopted the
lower value of $\Phi$, appropriate to black hole systems.  This may
reflect differences in the source spectrum as well as intrinsic
differences in $\Phi$ since we have assumed the same relation between
1.5--12$\;{\rm keV}$ count rate and total X-ray flux for all systems.
In addition to this effect, we note that the results for \Star{1705}
in Fig.~\ref{Fig:radiiratio} do not lie on a
\begin{figure}
  \epsfig{angle=0,width=84mm,file=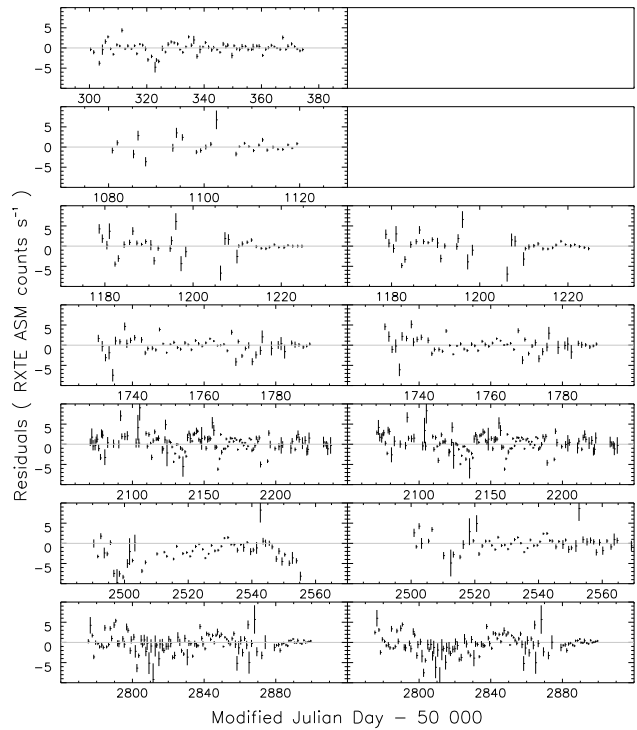}
  \caption{Residuals to the fitted portions of the {\it RXTE} ASM
  light curve of 4U~1705-44.  {\bf Left column:} using best-fitting
  parameters for each individual decay.  {\bf Right column:} Relating
  the brink luminosity to $\tau_e$ by $N_t/\tau_e=2.8\times10^{-5}$
  ASM counts ${\rm s}^{-2}$ in all cases.
  }
  \label{Fig:1705res}
\end{figure}%
straight line.

To examine this issue we first determined a best-fitting value
of $\Phi/\nu_{\rm{KR}}$ from decays {\it a} and {\it d}, which
we deem the strongest results by inspection of the light curve.  We
then demanded that this value of $\Phi/\nu_{\rm{KR}}$ apply to each
decay of \Star{1705}.  
In Fig.~\ref{Fig:1705res} we show the effect this has on the fits,
plotting the residuals for the individual best-fits, and for the
constrained fits.  If the original fits were strongly determined by
the data, we would expect the residuals to appear significantly worse
in the constrained fits.  In contrast, we find that in some
cases the residuals appear to improve, and in none is the fit
substantially worse; the reduced $\chi^2$ for the whole set of fits
increases from 11.4 when $L_t$ and $\tau_e$ are optimized individually
to 12.1 for the constrained fitting.  We note that technique is only
applicable when comparing decays of the same source, where variation
in the quantity $\Phi/\nu_{\rm{KR}}$ is minimized.

The large value of $R_{\rm{disc}}\!\left(L_t\right)$ for \Star{1751}
may be again the result of spectral differences, or in this case may
be the result of an erroneous source distance estimate.  \Star{1751}
has been assumed to lie at a distance of $8{\rm\,kpc}$, being possibly
associated with the Galactic Centre \citep[e.g.][]{GierlPouta05}; if
the true distance is more like $3{\rm\,kpc}$ then the two values of
$R_{\rm disc}$ will be approximately equal.  

The increasing value of $R_{\rm{disc}}\!\left(\tau_e\right)$ for the
2002 decay of \Star{1808} and the erratically increasing luminosity
between days 65 and 70 of Fig.~\ref{Fig:1808_2002} are both consistent
with the interpretation that the outer disc edge was not in the hot
state at the beginning of the decay.  As such, the disc radius
measured corresponds at first to a hot disc smaller than the disc as a
whole, which is therefore depleted on a shorter timescale.  At some
point the outer disc is heated and enters the hot state.  As well as
increasing the timescale of the decay, this introduces extra material
into the hot disc, causing rebrightening.

\section{Modified exponential limit}\label{Sec:dM2small}
We have assumed in the initial model that the limit of the exponential
decay is equal to the assumed constant mass transfer rate from the
donor star.  This requires that the hot disc is capable of supporting
the additional mass flux $-\dot M_2$ at all times and radii.  However,
the exponential nature of the decay itself implies that as the hot
disc is depleted in surface density, its mass transfer rate falls.
The mass transfer equation of the hot disc is linear in surface
density, which can therefore be decomposed into two components; one
representing the decay analogous to an unfed disc, the other
corresponding to the effects of feeding the hot disc at the rate
$-\dot M_2$.  The former component we denote $\Sigma_e$ and the latter
$\Sigma_2$.  Because the viscosity of the accretion disc depends on
temperature, and the temperature is declining during the decay, it is
necessary for $\Sigma_2$ to increase with time if it is to sustain the
same mass flux, implying that the corresponding flux $\mu_2$ must
decrease inwards.  Thus the limit of the central mass accretion rate
does not correspond to $-\dot M_2$, but rather $\mu_2\!(R_{\rm in})$.
The relevance of this issue is made clear by the light curves of
\Star{1705}, in which the limit of the exponential decay is often
below the long term mean count rate of $~12$ ASM counts ${\rm s}^{-1}$.

The standard derivation of the diffusion equation
\citep[e.g.][]{Pringle81} can be used to give the mass transfer rate
inward through the disc $\mu$ as
\begin{equation}
  \mu=6\pi R^{1/2}\left(\nu\Sigma R^{1/2}\right)'.
  \label{Eq:mu}
\end{equation}
Using the relation
\begin{equation}
  \mu'=2\pi R\dot\Sigma,
\end{equation}
we derive equation 9 of \citet{King98} as expected.  We require the
form given in (\ref{Eq:mu}) to prevent a constant of
integration arising.  

To make use of (\ref{Eq:mu}) we must adopt an appropriate form
for the viscosity $\nu$.  We use the $\alpha$-viscosity form
\begin{equation}
  \nu=\alpha c_sH,
\end{equation}
where $c_s$ is the sound speed, and the scale height is given by
equation 2 of \citet{King98};
\begin{equation}
  H=c_s\left(\frac{R^3}{GM_1}\right)^{\!1/2}.
\end{equation}
Using (\ref{Eq:TofR}) to substitute for $T$ and assuming that
$H\!(R)$ is a power law we obtain
\begin{equation}
  H=\left(\frac{k}{GM_1m}\right)^{\!4/7}\left(\frac{\left(1-\mathcal{A}\right)\left(n-1\right)}{4\pi\sigma}\right)^{\!1/7}
    L_\textrm{\sc x}^{1/7}R^{9/7},
  \label{Eq:Hpower}
\end{equation}
for a mean particle mass $m\simeq m_p$.  We write (\ref{Eq:Hpower}) in
terms of constant $H_0$ as
\begin{equation}
  H=H_0R_{\rm disc}\left(\frac{L_\textrm{\sc x}}{L_t}\right)^{\!1/7}\left(\frac{R}{R_{\rm disc}}\right)^{\!n},
\end{equation}
where $n=9/7$ is derived and $H_0\simeq0.2$ is adopted from KR.
\citet{BurdeKingSzusz98} note that if the X-ray flux is assumed to be
absorbed at the mid-plane, the power law index is changed to $45/38$;
we therefore consider values of $45/38\le n\le9/7$ as valid
to simplify subsequent derivation.  Using this form for $H$ in
(\ref{Eq:TofR}) to derive the viscosity we find that
\begin{equation}
  \nu=\nu_0\left(\frac{L_\textrm{\sc x}}{L_t}\right)^{\!2/7}\left(\frac{R}{R_{\rm disc}}\right)^{\!(9n-3)/8},
  \label{Eq:nu}
\end{equation}
where $\nu_0$ is defined as
\begin{equation}
  \nu_0=\alpha\left(\frac{k^4L_t\left(1-\mathcal{A}\right)\left(n-1\right)}{4\pi\sigma m^4}\right)^{\!1/8}
  H_0^{9/8}R_{\rm disc}^{3/4}.
\end{equation}
We simplify (\ref{Eq:nu}) by adopting $n=11/9$, so (\ref{Eq:mu})
becomes
\begin{equation}
  \mu=\frac{6\pi\nu_0}{R_{\rm disc}}\left(\frac{L_\textrm{\sc x}}{L_t}\right)^{\!2/7}R^{1/2}\left(\Sigma R^{3/2}\right)'.
\end{equation}
If we impose $\mu=-\dot M_2$ for all radii we obtain
\begin{equation}
  \Sigma_2=\frac{-\dot M_2}{3\pi\nu_0}\left(\frac{L_\textrm{\sc x}}{L_t}\right)^{\!-2/7}\frac{R_{\rm disc}}{R},
  \label{Eq:Sigma2}
\end{equation}
corresponding to a total contribution to mass in the hot disc of
\begin{equation}
  m_2=\frac{-2\dot M_2}{3\nu_0}\left(\frac{L_\textrm{\sc x}}{L_t}\right)^{\!-2/7}R_{\rm disc}^2,
\end{equation}
where the radius of the inner disc edge is much smaller than
$R_{\rm{disc}}$.  As $L_\textrm{\sc x}$ decreases, a greater surface
density and total mass is required to transport the same $-\dot M_2$
inward, so some of this material must be absorbed to increase the
surface density.  For a typical luminosity at the beginning of the
exponential phase 2.5 times greater than $L_t$ the change in mass
required is
\begin{equation}
  \frac{\Delta m_2}{\dot M_2}=\frac{2R_{\rm disc}^2}{3\nu_0}\left(1-2.5^{-2/7}\right).
  \label{Eq:deltam}
\end{equation}
Examining \Star{1705}, for which we have the largest mean luminosity
relative to $L_t$, we find a mean long term count rate of
$12{\rm\,ASM\,counts\,s}^{-1}$, whereas the limit of the decay {\it a}
for instance is $8.2{\rm\,ASM\,counts\,s}^{-1}$.  Taken over an
exponential phase duration of 57 days this makes the l.h.s. of
(\ref{Eq:deltam}) equivalent to $1.9\times10^7$ ASM units of mass.
Using the conversion factor from WebPIMMS of
$1{\rm\,ASM\,count\,s}^{-1}=$
$4\times10^{-13}{\rm\,J\,s}^{-1}{\rm\,m}^{-2}$, at a distance of
$7.3{\rm\,kpc}$ this integrated flux is equivalent to
$3\times10^{20}{\rm\,kg}$.  Using
$\nu_{\rm{KR}}=4\times10^{10}{\rm\,m}^{-2}{\rm\,s}^{-1}$ (as in
Table~\ref{Tab:calcpar}) we find
\begin{equation}
  R_{\rm disc}\simeq6.4\times10^8{\rm\,m},
\end{equation}
in strong agreement with the values given in Table~\ref{Tab:calcpar}
despite the weakness of the derivation.  In principle it is necessary
to simultaneously derive the time and radius dependencies of $\mu$
given the outer boundary condition of 
$\mu\!\left(R_{\rm disc}\right)=-\dot M_2$, but this is beyond the
scope of this paper.

\section{Conclusions}
A knee in the light curve of the decay from outburst of an transient LMXB is a
natural consequence of mass transfer onto the outer edge of the disc,
since this supply is effectively cut off from the compact object when
the outer disc enters the cool low-viscosity state.  When the knee can
be interpreted in this way we refer to it as a brink, since knees of
other types may be seen in the decay outburst.  The X-ray luminosity
at which the brink occurs is that at which the outer disc edge is just
kept hot by central illumination, allowing this radius to be
calculated.  In addition, the exponential time-scale of the decay gives
a second measure of the disc radius; these two estimates are in good
agreement.  By examining systems with well constrained disc radii we
deduced values of the constants $\Phi$ and $\nu_{\rm{KR}}$ required to
perform the calculations.  This allows more accurate results to be
obtained in other systems.  Where the source distance is poorly
constrained, the timescale $\tau_e$ allows the absolute luminosity
$L_t$ to be estimated, providing a measure of source distance.

As well as `secondary maxima' associated with the brink, we have
identified maxima whose natural explanation is the discontinuous
increase in the radius of the hot disc during outburst.  This increase
arises from initially self-shadowed disc regions becoming irradiated.
This process can occur until the whole disc is in the hot state.
It is characteristic for the time-scale of the exponential decay to
increase with each such maximum, and it is expected that the limit of
each exponential decay will also increase.  Although we do not have
sufficient data to examine the matter in detail, it is probable that
some secondary `maxima' of this type will also involve monotonic
decay, and therefore probably will appear as a knee in the X-ray
light curve.

Current instrumentation is capable of measuring the light curves of
X-ray transients in M31, for example \citet{TrudoPriedCordo06} show
part of a 2004 July decay of \Star{0043}.  While this light curve,
with only 4 data points over 3 consecutive days, is too sparse to
allow us to deduce any parameters from fitting the decay, our method
could be applied to better-sampled extragalactic transient
decays. Since the distance is relatively well-known for objects in
external galaxies, we could use our method to deduce their accretion
disc radii.  This would contribute to constraints on the fundamental
system parameters, such as orbital separation and mass ratio, which
might otherwise be impossible to determine.

\section*{Acknowledgments}
We thank Andrew King for comments on an earlier version of this work.
CRP was supported by a PPARC studentship, and thanks the OU Dept. of
Physics and Astronomy for funding during the final write-up of this
paper. MF acknowledges the CNRS for financial support.
This research has made use of data obtained through the High
Energy Astrophysics Science Archive Research Center Online Service,
provided by NASA's Goddard Space Flight Center (GSFC). ASM results
were provided by the ASM and RXTE teams at MIT and at the RXTE SOF and
GOF at NASA's GSFC.

\label{lastpage}


\begin{thebibliography}{99}
\bibitem[\protect\citeauthoryear{Burderi, King \& Szuszkiewicz}{1998}]{BurdeKingSzusz98}
  Burderi L., King A.~R., Szuszkiewicz E.,
  1998, ApJ, 509, 85
\bibitem[\protect\citeauthoryear{Campana et al.}{2003}]{CampanaETAL03}
  Campana S., Ravasio M., Israel G.~L., Mangano V., Belloni T.,
  2003, ApJ, 594, L39
\bibitem[\protect\citeauthoryear{Chakrabarty \& Morgan}{1998}]{ChakrMorgan98}
  Chakrabarty D., Morgan E.~H.,
  1998, Nat, 394, 346
\bibitem[\protect\citeauthoryear{Falanga et al.}{2005}]{FalangaETAL05}
  Falanga M. et al.,
  2005, A\&A, 436, 647
\bibitem[\protect\citeauthoryear{Foulkes, Haswell \& Murray}{2006}]{FoulkesETAL06}
  Foulkes S.~B., Haswell C.~A., Murray J.~R.,
  2006, MNRAS, 366, 1399
\bibitem[\protect\citeauthoryear{Galloway et al.}{2002}]{GallowayETAL02}
  Galloway D.~K., Chakrabarty D., Morgan E.~H., Remillard R.~A.,
  2002, ApJ, 576, L137
\bibitem[\protect\citeauthoryear{Gierli{\'n}ski \& Poutanen}{2005}]{GierlPouta05}
  Gierli{\'n}ski M., Poutanen J.,
  2005, MNRAS, 359, 1261
\bibitem[\protect\citeauthoryear{Haberl \& Titarchuk}{1995}]{HaberlTitar95}
  Haberl F., Titarchuk L.,
  1995, A\&A, 299, 414
\bibitem[\protect\citeauthoryear{Hynes et al.}{1998}]{HynesETAL98}
  Hynes R.~I. et al.
  1998, MNRAS, 300, 64
\bibitem[\protect\citeauthoryear{Hynes et al.}{2004}]{HynesETAL04}
  Hynes R.~I., Steeghs D., Casares J.,Charles P.~A., O'Brien K.,
  2004, ApJ, 609, 317
\bibitem[\protect\citeauthoryear{de Jong, van Paradijs \& Augusteijn}{1996}]{JongParadAugus96}
  de Jong J.~A., van Paradijs J., Augusteijn T.,
  1996, A\&A, 314, 484
\bibitem[\protect\citeauthoryear{King \& Ritter}{1998}]{KingRitter98}
  King A.~R., Ritter H.,
  1998, MNRAS, 293, L42 (KR)
\bibitem[\protect\citeauthoryear{King}{1998}]{King98}
  King A.~R.,
  1998, MNRAS, 296, L45
\bibitem[\protect\citeauthoryear{Markwardt et al.}{2002}]{MarkwardtETAL02}
  Markwardt C.~B., Swank J.~H., Strohmayer T.~E., in~'t Zand J.~J., Marshall F.~E.
  2002, ApJ, 575, L21
\bibitem[\protect\citeauthoryear{Nishiuchi et al.}{1999}]{NishiuchiETAL99}
  Nishiuchi M. et al.,
  1999, ApJ, 517, 436
\bibitem[\protect\citeauthoryear{Orosz \& Bailyn}{1997}]{OroszBailyn97}
  Orosz J.~A., Bailyn C.~D.,
  1997, ApJ, 477, 876
\bibitem[\protect\citeauthoryear{Orosz et al.}{1998}]{OroszETAL98}
  Orosz J.~A., Jain R.~K., Bailyn C.~D., McClintock J.~E., Remillard R.~A.,
  1998, ApJ, 499, 375
\bibitem[\protect\citeauthoryear{Orosz et al.}{2002}]{OroszETAL02}
  Orosz J.~A. et al.,
  2002, ApJ, 568, 845
\bibitem[\protect\citeauthoryear{Park et al.}{2004}]{ParkETAL04}
  Park S.~Q. et al.,
  2004, ApJ, 610, 378
\bibitem[\protect\citeauthoryear{Pringle}{1981}]{Pringle81}
  Pringle J.~E.,
  1981, ARA\&A, 19, 137
\bibitem[\protect\citeauthoryear{Pringle}{1986}]{Pringle96}
  Pringle J.~E.,
  1996, MNRAS, 281, 357
\bibitem[\protect\citeauthoryear{Rappaport \& Joss}{1997}]{RappaJoss97}
  Rappaport S., Joss P.~C.,
  1997, ApJ, 486, 435
\bibitem[\protect\citeauthoryear{Shahbaz, Charles \& King}{1998}]{ShahbCharlKing98}
  Shahbaz T., Charles P.~A., King A.~R.,
  1998, MNRAS, 301, 382 (SCK)
\bibitem[\protect\citeauthoryear{Trudolyubov, Priedhorsky \& Cordova}{2006}]{TrudoPriedCordo06}
  Trudolyubov S., Priedhorsky W., Cordova F.,
  2006, ApJ, 645, 277
\bibitem[\protect\citeauthoryear{in 't Zand et al.}{2001}]{ZandETAL01}
  in 't Zand J.~J.~M. et al.,
  2001, A\&A, 372, 916
\end{thebibliography}
\end{document}